# Density and Glass Forming Ability in Amorphous Atomic Alloys: the Role of the Particle Softness

Ian Douglass, Toby Hudson and Peter Harrowell

School of Chemistry, University of Sydney

Abstract

A key property of glass forming alloys, the anomalously small volume difference with respect to the crystal, is shown to arise as a direct consequence of the soft repulsive potentials between metals. This feature of the inter-atomic potential is demonstrated to be responsible for a significant component of the glass forming ability of alloys due to the decrease in the enthalpy of fusion and the associated depression of the freezing point.

## 1. Introduction

The density difference between a liquid and the crystal phases represents a fundamental thermodynamic feature of freezing. When observed, the increased density of the crystal underscores the improved packing afforded by crystalline arrangements of atoms. For example, in the common model of a hard sphere liquid, the appearance of the ordered crystal structure can be understood completely in terms of the entropy gain achieved through this packing improvement [1]. In this model, the fractional molar volume difference $\delta = (V_{liq} - V_{xtl})/V_{xtl}$ is 0.156 [2]. In glass forming metal alloys, however, $\delta$ is significantly less than the hard sphere value [3-5]. In Figure 1 we have collected experimental values of the fractional volume difference for pure metals and alloys. For mixtures, the crystal volume is taken to be $V_{xtl} = \sum_i x_i V_{xtl}^i$ where $x_i$ is the molar fraction of species i and $V_{xtl}^i$ is the molar volume of the pure crystal of species i. This choice of $V_{xtl}$ is convenient for a general survey since, unlike the crystal volumes for specific compounds, it allows a crystal volume to be assigned for arbitrary compositions and makes use only of the easily obtained volumes of the pure elements. When we come to consider crystallization kinetics we shall use the volume of the specific crystal formed. While the inert gases exhibit a similar fractional volume change to that found for the hard sphere model, the value of $\delta$ for pure metals is smaller by a factor of ~ 3. In the case of bulk metallic glasses (alloys of between 3 to 5 components), $\delta$ is even smaller with many examples where the volume of the amorphous alloy is equal to (i.e. $\delta=0$) or less (i.e. $\delta < 0$) than $V_{xtl}$. Even when accounting for the improved packing ability of binary mixtures with increasing radius ratio $\gamma = r_{large}/r_{small}$, the value of $\delta$ obtained for dense packed hard sphere crystals and glasses never drops below 0.1.

From Figure 1, it is clear that the treatment of the amorphous state as a random close packing of those spheres cannot account for the values of $\delta$ observed experimentally in metal alloys. This failure is noteworthy. The hard sphere model has long been regarded as a realistic model of atomic liquid structure [7], so understanding the physical origin of this discrepancy is important. Furthermore, a correlation has been noted between the value of $\delta$ and the glass forming ability of metal alloys [8,9]. This correlation is demonstrated in Fig.2 where the critical cooling rate of a range of metallic glasses is shown to increase significantly with increasing $\delta$.

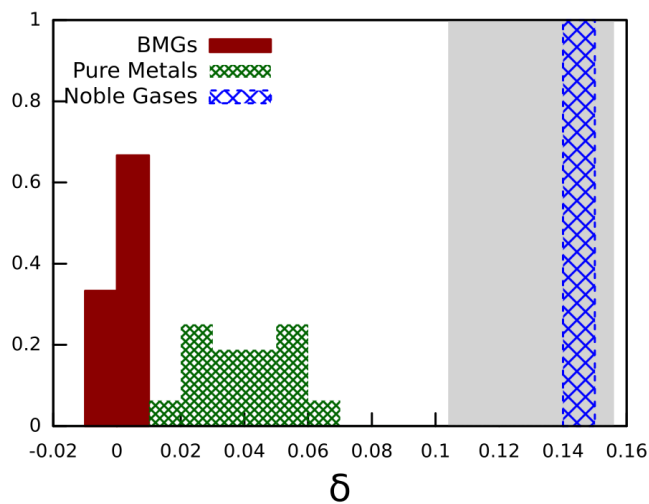

**Figure 1**. The distribution of the fractional volume difference δ among noble gases [6], pure metals [6] and bulk metallic glasses (BMG's) [5]. The range of δ exhibited by binary hard sphere mixtures [2] with radius ratio γ in the range $1 \leq \gamma \leq 1.76$ is indicated by the gray shaded region.

A number of researchers have proposed that the anomalously high density of the bulk metallic glasses is the result of deviations from the random close packing model in the form of chemically ordered clusters. This idea dates back to the 1981 paper by Gaskell [10] in which he argued that the local structure of the glassy alloy was similar to that found in the compound crystal. Miracle [11] has extended this argument, proposing that metallic glasses be described as periodic arrays of solute centred clusters. To date, it has not been explicitly demonstrated that any of these alternate structures can account for amorphous packings of the observed density.

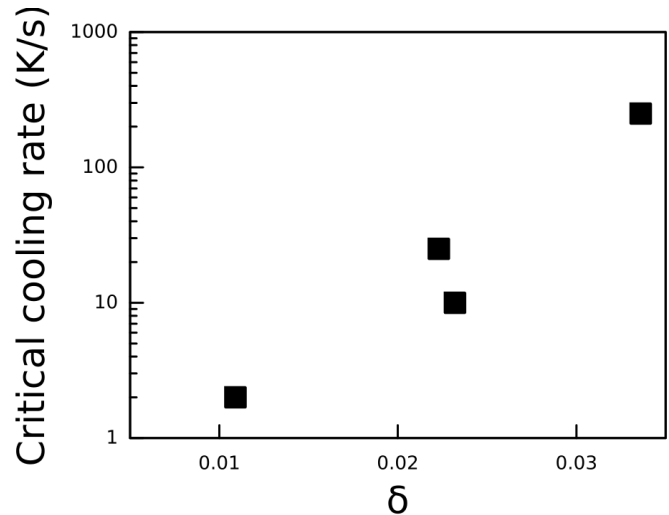

**Figure 2.** The critical cooling rate for Zr-based glass forming alloys as a function of δ using data from ref. 12.

The alternative conclusion to draw from the data in Fig. 1 is that it is the hard sphere model, rather than the assumption of random close packing, that is at fault. Egami [13] has made this point, noting that the low values of the Gruneisen parameter and the ratio $T_m/T_b$ of the freezing point and the boiling point in metal alloys indicate that the atomic repulsions must be significantly softer than the Lennard-Jones potentials developed for the inert gases. While the influence of the potential softness on the value of δ does not appear to have been systematically studied, accurate atomic potentials based on the Embedded Atom Method [14] (EAM) – which employ 'soft' repulsions [15,16] of the Born-Mayer form,

$\exp\left[-2\alpha\left(\frac{r_{ij}}{\sigma_{ij}}-1\right)\right]$, do successfully reproduce the observed (small) values of δ.

In this paper we shall explore the proposition that the essential feature required to account for the small or negative values of the fractional volume change is a short range repulsion significantly softer than that of hard spheres or the $r^{-12}$ repulsion commonly used to model the inert gas atoms.

## 2. Models and Methodology

For the task of studying the role of particle softness we have chosen the Morse potential $u_{ij}(r)$ between species i and species j,

$$u_{ij}(r) = \varepsilon\left(\exp\left[-2\alpha_{ij}\left(\frac{r}{\sigma_{ij}}-1\right)\right] - 2\exp\left[-\alpha_{ij}\left(\frac{r}{\sigma_{ij}}-1\right)\right]\right) \quad (1)$$

where $\sigma_{ij}$ and $\alpha_{ij}$ set the interparticle length scale and potential softness. In this study we shall only consider the case where $\sigma_{ij} = (\sigma_{ii} + \sigma_{jj})/2$ and $\alpha_{ij} = \alpha$ for all *ij* pairs. The size ratio γ is defined as $\gamma = \sigma_{AA}/\sigma_{BB}$. The repulsion term in the Morse potential is the same as that used in the more accurate EAM potentials. The Morse potential has previously been used to model transition metal interactions [17] with the reported empirical values of α in the range $3.89 \leq \alpha \leq 4.33$. For comparison, the $r^{-12}$ repulsion used in the Lennard-Jones potential is best fitted with α = 6 (see Fig. 3). Molecular dynamics were carried out using the GROMACS suite [18] with ~10,000 atoms simulated in an NPT ensemble with pressure fixed to zero for all simulations, and using Berendsen-style thermo- and barostats [18]. The Morse potential was truncated to 2.5σ using a force-shift method, and adjusted such that the potential minima of interaction was constrained to the same separation for all values of α.

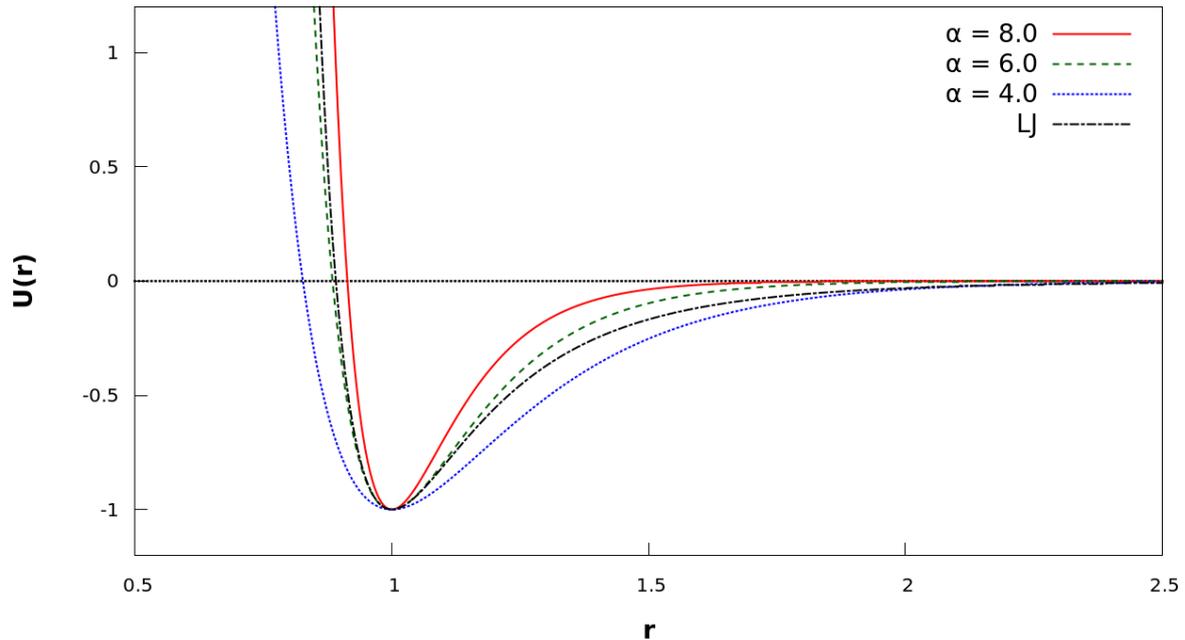

**Figure 3**. Morse potentials for a variety of values of the softness parameter α. The Lennard-Jones (LJ) potential is included for comparison.

We have considered three different alloys, distinguished by the size ratio and composition: $\gamma = 1.224$ with $AB_2$; $\gamma = 1.366$ with AB; and $\gamma = 1.543$ with $AB_3$, where $\gamma = \sigma_{AA}/\sigma_{BB}$. These size ratios and compositions were selected as they each are associated with a close packed compound crystal [19]. Only one alloy, $\gamma = 1.224$ with $AB_2$, was observed to crystallize and we examine the kinetics of this transition in this paper.

The analysis we shall present requires that we take into account the dependence of the size of atoms on the softness parameter α. In this paper we shall define the radius $r_i$ of an atomic species $i$ at a given temperature, pressure and choice of α as half the distance to the position of the first peak of the radial distribution function between two particles of species $i$.

## 3.1 Fractional Volume Change and the Phase-Dependence of Atomic Size

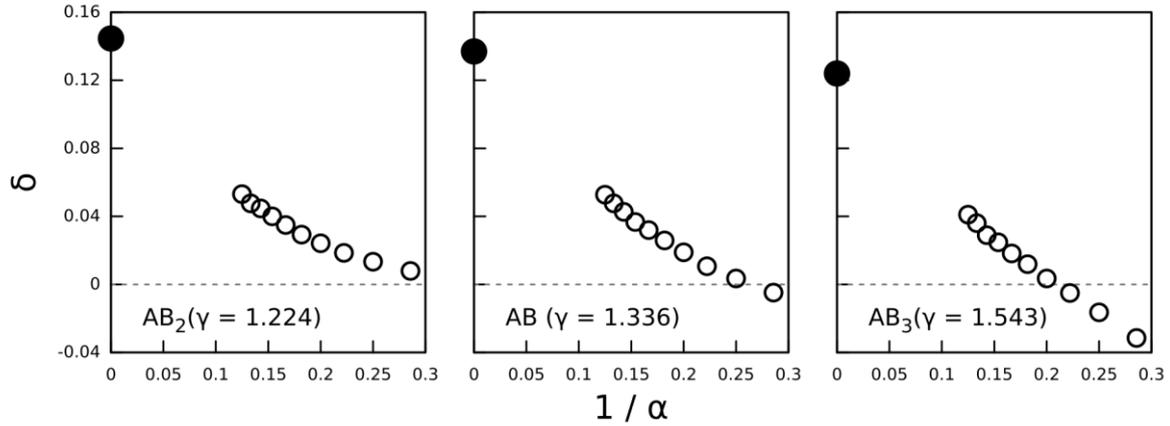

**Figure 4.** The fractional volume change δ (open circles) as a function of 1/α for the three Morse alloys as indicated calculated at T = 0.2. 1/α = 0 corresponds to the hard sphere result (filled circles) where we have used the close packed face centred crystal and the random closed packed density for the binary mixtures calculated as in ref. 2.

In Fig. 4 we plot the fractional volume change $\delta = (V_{liq} - V_{xtl})/V_{xtl}$ as a function of the value of α for three different choices of the size ratio γ. The reader is reminded that for the calculation of δ we define the crystal volume $V_{xtl}$ as the composition average of the single component (face centered cubic) crystals (see Section 1). As the particle softness (i.e 1/α) increases we find a systematic decrease in the fractional volume change δ, reaching values of ~ zero or negative for α = 4, a value typically used to model inter-metallic repulsions. We conclude that the small density difference between liquid and crystal is a result of the soft repulsions characteristic of metals.

How does particle softness produce these high liquid densities, relative to the crystal? The suggestion that it is a result of improved particle packing can be directly tested. The values of atomic radii $r_i$ were obtained from the position of the first peak in $g_{ii}(r)$ the liquid pair distribution function between like species as discussed in the previous Section. The efficiency of the packing can then be determined using the packing fraction $\eta$ where $\eta = \frac{1}{V}\sum_i N_i \frac{4\pi r_i^3}{3}$ with the sum being over the different species. The packing fraction of each liquid has been calculated using the α-dependent radii and the results plotted against the potential softness in Fig. 5. We find that the liquid packing fraction exhibits a small (~ 5%) increase over the hard sphere value with decreasing α, the size of the effect depending on γ and composition. We find that the increase in packing fraction accounts for 20% to 50% of the total decrease in δ below the hard sphere value. We conclude that the decrease in δ demonstrated in Fig. 4 cannot be accounted for simply by the presence of especially dense local structures as suggested.

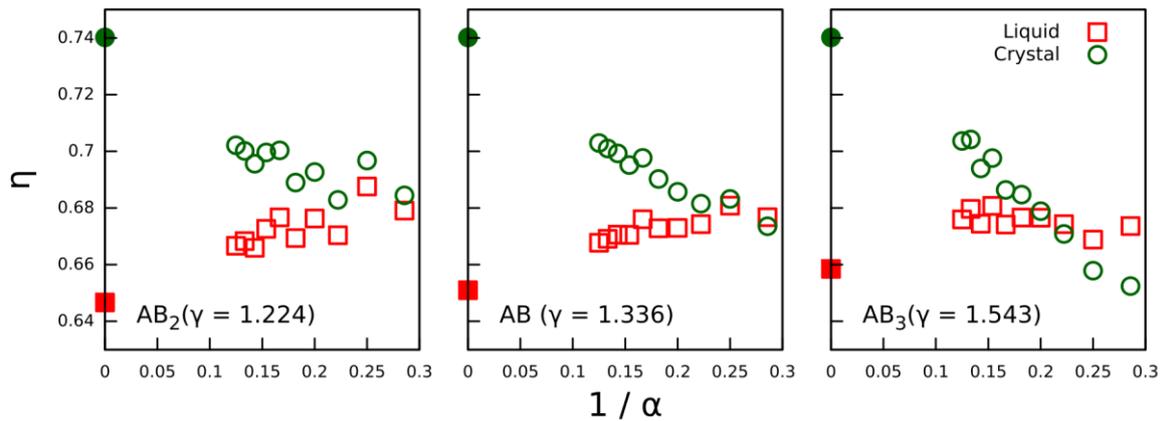

**Figure 5.** The packing fraction η for the liquid and the crystal (see text) as a function of 1/α for the three different alloys.

In Fig. 5 we have also plotted the packing fraction for the crystal. To calculate the relevant packing fraction for a crystal volume defined as the composition average of the two pure crystals we simply insert $V_{xtl}$, as defined in Section 1 using appropriately simulated crystals, into our expression for the packing fraction, again using the radius obtained from the liquid structure. In clear contrast to the liquid, we find that the crystal packing fractions *decrease* with increasing particle softness. What this means is that even when the crystal density is increasing as the particle softens, it does not increase as much as would have been predicted by the decrease in the particle size (as measured from the liquid structure). Apparently, the constraint of the crystal structure reduces the degree to which the condensed phase can exploit the increase in compressibility of the particles. It is this decrease in crystal packing fraction that makes up the remaining decrease in the packing fraction δ with decreasing α. Note that the vanishing of δ does not represent a special condition (e.g. an asymptote in 1/α) and, as shown in Fig. 4, the volume difference can become negative. In summary, the reduction of δ with increasing particle softness (i.e. decreasing α) arises because the increasing compression of the liquid (measured here as decreasing particle radii) exceeds the increase in the compression of the crystal with 1/α. As we increase the softness of the interaction, particles increasingly behave as if they are larger in the crystal than in the liquid.

### 3.2 Crystallization Kinetics and the Influence of Particle Softness

Next we consider the glass forming ability of the alloys and its correlation with the factional volume difference. Does the decrease in δ observed in the Morse alloys coincide in an improvement of glass forming ability? The alloy $AB_2$ (γ = 1.224) crystallizes into the Laves $MgZn_2$ structure, as seen in previous Lennard-Jones studies [20]. (Strictly, the simulation in ref. 20 used a size ratio γ = 1.2 rather than the value γ = 1.224, corresponding to ideal ratio when packing hard spheres into the $MgZn_2$ structure, used here.) This structure can be quantified in terms of common neighbour analysis. As shown in Fig. 6, a particular feature of the crystal is A-A pairs with 6 common neighbours, a feature we have referred to as a Frank-Kasper (FK) 'bond' [20]. In the crystal, each A atom is involved with a tetrahedral arrangement of 4 FK bonds.

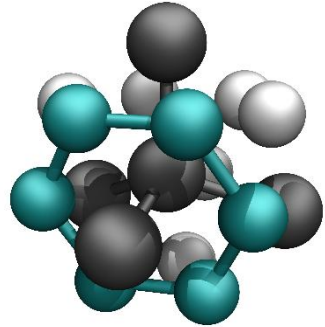

**Figure 6.** The structural resolution of the MgZn$_2$ crystal. A pair of particles (black) from an Frank-Kaper (FK) bond if they share a ring of 6 common neighbours (blue). In the MgZn$_2$ crystal these FK bonds are organised in a diamond-like structure as indicated so that a particle that is involved in four FK bonds is considered 'crystalline'.

The dependence of the kinetics of crystallization of the AB$_2$ γ=1.224 alloy into the MgZn$_2$ crystal structure on α was studied by direct MD simulation as follows. Liquids were equilibrated at T = 0.6 and then quenched to the final temperature over $10^5$ τ. After a further time of $10^6$ τ, the degree of crystallization was monitored by the concentration of 4FK particles. If the concentration exceeded 1% then the crystallization was deemed to have begun. In Fig. 7 we plot the results of this analysis. For the liquids with α = 8, 6 and 5 we find a range of temperatures at which crystallization was observed. The temperature range is bounded above because of the small thermodynamic driving 'force' and below because of the low particle mobility. At α=4 we cease to find any significant crystallization, indicating a marked increase in the glass forming ability.

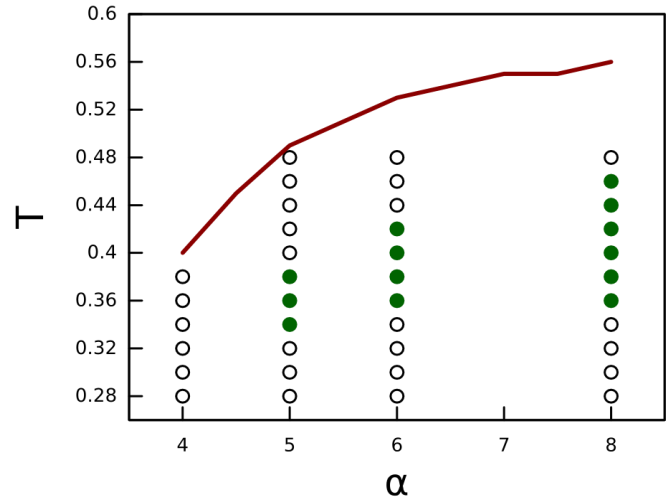

**Figure 7.** Observation of crystallization as a function of α. Filled circles indicate temperatures at which > 1% of the sample was crystalline after $10^6$ τ at the indicated final temperature. Empty circles indicate temperatures at which no crystallization was observed. The solid curve is the melting point $T_m$.

Since the rate of crystallization can depend on many factors, we must rely on a model calculation to identify exactly how particle softening achieves the observed slow down of crystallization. To this end we shall follow Turnbull [21] and combine the classical theories of crystal nucleation and growth to calculate the crystallization time τ as a function of temperature. To begin, if we assume a constant nucleation rate I, then [21]

$$\log(\tau) = \frac{1}{4}\log(X) - \frac{1}{4}(\log I + 3\log u) \qquad (2)$$

where $X = 10^{-4}$, an arbitrary threshold fraction of the volume that needs to be crystalline in order to assert that crystallization has occurred, and $u$ is the crystal growth rate. (Note all references to "log" are in base 10.) The nucleation rate I is given by

$$\log(I) = (\ln D - \frac{\Delta G^*}{T})/2.302 \qquad (3)$$

From classical nucleation theory we have

$$\Delta G^*/T = \frac{16\pi}{3} \frac{a^3 b T_m^3}{T(T-T_m)^2} \tag{4}$$

where

$$a = \frac{\sigma}{\rho_c^{2/3} \Delta h_m} \tag{5}$$

and

$$b = \frac{\Delta h_m}{T_m} \tag{6}$$

with $\Delta h_m$, the enthalpy per particle of fusion, $\sigma$ is the crystal-melt surface free energy, $T_m$ is the melting point and $\rho_c$ is the crystal density at $T_m$. Finally, using the Wilson-Frenkel expression [21] for the crystal growth rate, $u$, we write

$$\log(u) = \left(\ln D + \ln\left(1 - \exp\left[\frac{b(T_m - T)}{T}\right]\right)\right)/2.302 \tag{7}$$

Next, we must determine the values of the various parameters in the equations above. Atomic mobility in the supercooled liquid plays a central role in determining the crystallization rate as well as testing propositions concerning the dynamic influence of liquid density. A plot of the diffusion constant vs T for the larger atomic species in the AB$_2$ ($\gamma = 1.224$) liquid alloy for four different values of α is shown in Fig. 8. At low temperature, we find that the diffusion constant increases as the potentials soften, in agreement with previous reports [22]. (The fact that mobility increases even as the volume difference between crystal and liquid all but vanishes, provides an explicit rebuttal of the free volume argument which would predict that as δ approaches zero the diffusion constant should *decrease* significantly.) The calculations of the crystallization times require extrapolation of the diffusion data to lower temperatures.

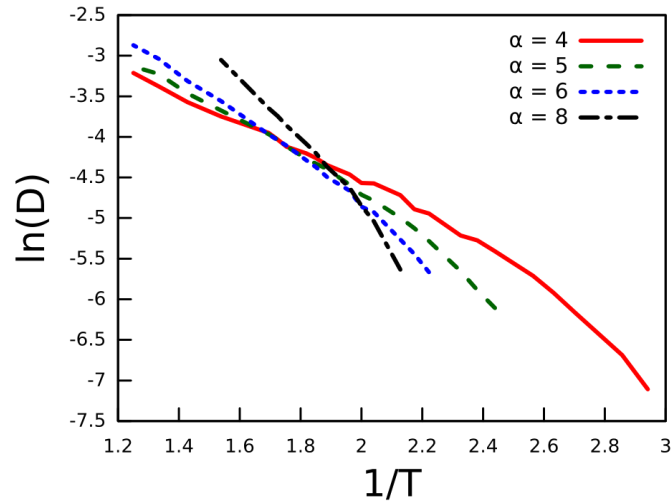

**Figure 8.** The large particle self diffusion constant D in the $AB_2$ ($\gamma = 1.224$) liquid alloy as a function of inverse temperature for a number of values of $\alpha$.

The temperature dependence of the self diffusion constant for the large particles in the $AB_2$ ($\gamma=1.224$) liquid mixture has been fitted to the VFT expression [23],

$$D = \nu \exp\left(-\frac{E_a}{T-T_o}\right) \qquad (8)$$

The fit parameters are provided in Table 1.

| α | $E_a$ | ν | $T_o$ |
|---|---|---|---|
| 4 | 0.7549 | 0.12324 | 0.1861 |
| 5 | 0.8829 | 0.19443 | 0.2106 |
| 6 | 1.1875 | 0.39956 | 0.1975 |
| 8 | 0.9830 | 0.62986 | 0.2775 |

**Table 1** Fit parameters for the VFT expression (Eq. 8) for the temperature dependence of the diffusion constant.

Next, we must determine the melting point $T_m$. The alloy $AB_2$ $\gamma = 1.224$ crystallizes spontaneously into the Laves structure $MgZn_2$ described in ref. 20. This structure can be quantified in terms of common neighbour analysis. A particular feature of the crystal is A-A pairs with 6 common neighbours, a feature we have referred to as a Frank-Kasper (FK) 'bond' [20]. In the crystal, each A atom is involved with a tetrahedral arrangement of 4 FK bonds. To determine the melting point of the crystal for different values of α we have carried out the following calculations. An initial condition was established consisting of a slab of crystal with liquid on either side by melting a crystal while constraining the atoms in the crystal slab region. From this starting point the system was quenched to the temperature of interest and observed after 500,000 time steps. The presence of both phases in the initial conditions ensures that metastability was not a problem. The potential energy, diffusion constant and number of A atoms with 4 FK bonds were measured to monitor the transition between crystal and liquid. In Fig. 9 we plot the number of A atoms with 4 FK bonds as a function of T

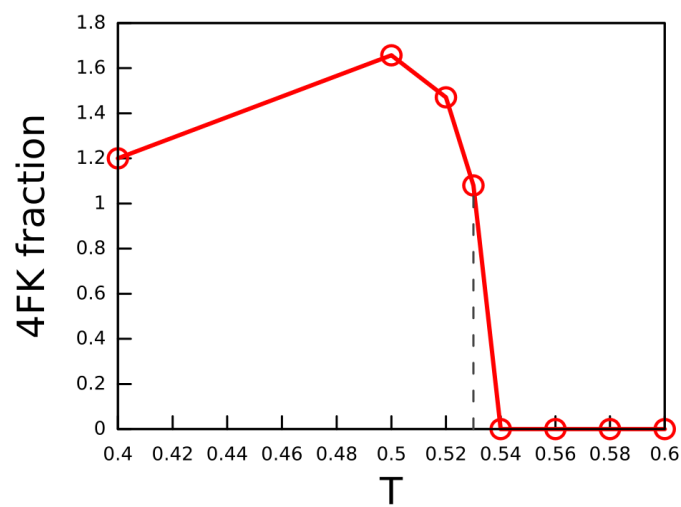

**Figure 9.** The fraction of 4FK particles (as described in the text) relative to the initial number of 4FK particles after 500,000 time steps at the indicated T from a starting configuration consisting of a planar slab of crystal in contact with the liquid phase for the $AB_2$ ($\gamma$=1.224) mixture with $\alpha$= 6.0. An increase or decrease in the number of 4FK bonds corresponds to freezing and melting, respectively. The melting point $T_m$ is indicated by the value of T at which no change in the amount of order was observed.

To determine the interfacial free energy $\sigma$ we have carried out the following calculations. Instead of the planar crystal slab, we have set up an initial configuration with a spherical cluster of crystal surrounded by a liquid using the following protocol:

i) Starting from a crystal configuration equilibrated to the final, particle with a spherical volume of radius r were identified and their positions constrained.

ii) The system was heated to above the melting point resulting in a liquid surrounding the constrained crystal cluster of radius r.

iii) The system was the cooled to the melting point and allowed to equilibrate before being quenched to supercooled temperature at which point the constraints on the crystal particles was removed.

iv) The potential energy was monitored as a function of time (see Fig. 10). A persistent decrease in the potential energy indicates crystal growth and, hence, establishes that the cluster was super-critical while an increase in potential energy with time signalled the melting of the cluster and so indicated a sub-critical cluster. By interpolating between the radius of the smallest growing cluster and largest melting cluster we determine the critical radius $r_c$. Values of $r_c$ for different $\alpha$'s are listed in Table 2. The temperature in each case was adjusted to give a constant chemical potential difference between crystal and liquid of -0.1187.

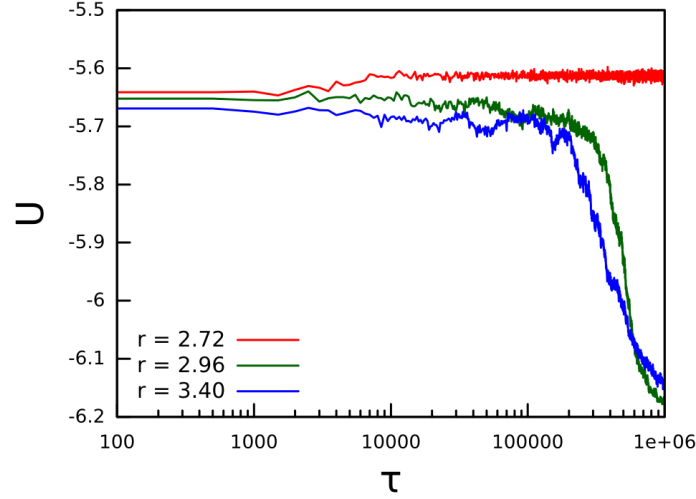

**Figure 10.** The time evolution of the potential energy U during runs with $\alpha = 5$ starting with a crystal cluster of radius r, as indicated, at a supercooled temperature of 0.425. A decrease in U indicates that crystallization has taken place while a (small) increase in U is a result of the melting of the initial crystal cluster. In this example, $r_c$ must lie between 2.72 and 2.96.

The surface free energy $\sigma$ was extracted from the critical radius $r_c$ by applying the result from classical nucleation theory, i.e.

$$\sigma = -\frac{r_c \rho_c \Delta\mu}{2} \qquad (9)$$

where $\rho_c$ is the density of the crystal at the supercooled temperature and the chemical potential difference $\Delta\mu$ is estimated by the following expression,

$$\Delta\mu = \frac{\Delta h_m}{T_m}(T_m - T) \qquad (10)$$

where $\Delta h_m$ is the heat of fusion per particle.

| α | Δh$_m$ | σ | T$_m$ | ρ$_c$ | r$_c$ |
|---|---|---|---|---|---|
| 4 | -0.5934 | 0.3368 | 0.40 | 2.5590 | 2.19 |
| 5 | -0.8986 | 0.3576 | 0.49 | 2.1612 | 3.51 |
| 6 | -1.1580 | 0.4589 | 0.53 | 2.0265 | 5.29 |
| 8 | -1.6745 | 0.5531 | 0.56 | 1.9473 | 6.46 |

**Table 2.** The heat of fusion/particle Δh$_m$, crystal-liquid surface free energy σ, melting point T$_m$, crystal density ρ$_c$ and the critical nucleus radius r$_c$ for the AB$_2$ ($\gamma = 1.224$) alloy for four different values of α.

The calculated crystallization times as a function of temperature for the $\gamma = 1.224$ alloy have been plotted for 4 values of α in Fig. 11. Note that in these calculations we use the VFT expression for the diffusion constants at temperatures below the temperature range at which the VFT parameters (Table 1) were fitted.

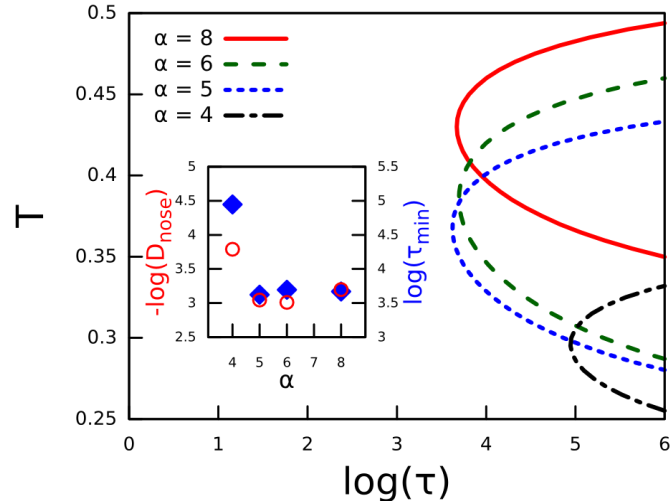

**Figure 11.** The crystallization time $\tau$ as a function of temperature for the $AB_2$ ($\gamma = 1.224$) alloy for four different values of α. The larger the minimum crystallization time, the greater the glass forming ability of the alloy. Insert: The minimum crystallization time $\tau_{min}$ (i.e. the 'nose' of the TTT curve) exhibits a dramatic increase as α decreases from 5 to 4 in qualitative agreement with the observations from MD simulations (see Fig. 6). The increase in the liquid relaxation time $D^{-1}$ at T corresponding to $\tau_{min}$ exhibits a similar sharp increase with decreasing α.

As shown in the insert of Fig. 11, the minimum crystallization time $\tau_{min}$ exhibits an abrupt increase as α decreases from 5 to 4, in qualitative agreement with our observation in Fig. 6 that crystallization ceases to be observed on quenching at this same decrease in α. Having reproduced the observed trend of glass forming ability with the calculated TTT curves, we can now resolve the origin of this trend (at least within the limits of the classical model). At the temperature $T_{nose}$ corresponding to the minimum crystallization time (i.e the 'nose' of the TTT curve), we find that the dominant contribution (~55%) to the increase in $\tau_{min}$ at α=4 is the increase in the liquid relaxation time, $D^{-1}$ (see Fig. 11). As $T_{nose} \sim 0.75 T_m$ for all values of α, this slow down of the liquid can be traced to the marked decrease in the melting point $T_m$ as α decreases. ($T_m$ = 0.56, 0.53, 0.49 and 0.40 for α = 8, 6, 5 and 4, respectively.) As the liquid softens, the decreasing melting point means that crystallization requires temperatures low enough that, even with the increased mobility due to softening, the liquid dynamics is considerably slower than that found in the stiffer potentials at their respective $T_{nose}$'s. This depression of the freezing point is a result of the decrease in the enthalpy of fusion which, in turn, can be attributed to the decrease in the density difference between crystal and liquid. The decrease in the enthalpy of fusion is also responsible for the decrease in the thermodynamic driving of crystal growth and the increase in the reduced surface tension, $\sigma/(\rho^{2/3}\Delta h_m)$, the other contributors to the slowdown in crystallization.

## 4. Conclusions

In this paper we have presented evidence based on simulations of binary alloys modelled with Morse potentials to demonstrate that the small fractional volume difference between the liquid and crystal phases of metal alloys can be largely attributed to the soft repulsive potentials. Previous reports in which the high relative density of the liquid was attributed to the presence of specific well packed local structures is not supported by our results, in which we show that the packing fraction varies little as δ is decreased. Instead, we find that the effective atomic radii in the amorphous state become increasingly smaller than those in the crystal as the softness of the particles increases.

We have demonstrated that small values of δ are indeed correlated with improved glass forming ability. This correlation is *not* because the small 'free volume' renders the amorphous mobility low, as previously suggested [8]. In fact, the softer potentials enhance mobility. We find that the improved glass forming ability is a consequence of the decrease in the enthalpy difference between the crystal and liquid as the particles become softer. Our results indicate that hard sphere-like models of alloys fail to reproduce important features of glass forming alloys and that the effects of softening the particle interactions can be significant and need to be included in accounting for the glass forming ability of alloys.

**Acknowledgements**

The authors gratefully acknowledge financial support for the Australian Research Council.